\documentclass[%
 reprint,
 superscriptaddress,
 amsmath,amssymb,
 prl,
]{revtex4-2}

\usepackage{graphicx,float}
\usepackage[caption=false]{subfig}
\usepackage{dcolumn}
\usepackage{bm}
\usepackage{hyperref}
\usepackage{sidecap}
\usepackage{siunitx}

\begin{document}

\title{Multi-GeV Wakefield Acceleration in a Plasma-Modulated Plasma Accelerator}

\author{J. J. van de Wetering}
\email{johannes.vandewetering@physics.ox.ac.uk}
\affiliation{John Adams Institute for Accelerator Science and Department of Physics, University of Oxford, Denys Wilkinson Building, Keble Road, Oxford OX1 3RH, United Kingdom}%
\author{S. M. Hooker}%
\affiliation{John Adams Institute for Accelerator Science and Department of Physics, University of Oxford, Denys Wilkinson Building, Keble Road, Oxford OX1 3RH, United Kingdom}%
\author{R. Walczak}%
\affiliation{John Adams Institute for Accelerator Science and Department of Physics, University of Oxford, Denys Wilkinson Building, Keble Road, Oxford OX1 3RH, United Kingdom}%
\affiliation{Somerville College, Woodstock Road, Oxford OX2 6HD, United  Kingdom}%

\date{\today}

\begin{abstract}
We investigate the accelerator stage of a Plasma-Modulated Plasma Accelerator (P-MoPA) [Phys. Rev. Lett. \textbf{127}, 184801 (2021)] using both the paraxial wave equation and particle-in-cell (PIC) simulations. We show that adjusting the laser and plasma parameters of the modulator stage of a P-MoPA allows the temporal profile of pulses within the pulse train to be controlled, which in turn allows the wake amplitude in the accelerator stage to be as much as $72\%$ larger than that generated by a plasma beat-wave accelerator with the same total drive laser energy. Our analysis shows that Rosenbluth-Liu detuning is unimportant in a P-MoPA if the number of pulses in the train is less than $\sim 30$, and that this detuning is also partially counteracted by increased red-shifting, and hence increased pulse spacing, towards the back of the train. An analysis of transverse mode oscillations of the driving pulse train is found to be in good agreement with 2D PIC simulations. PIC simulations demonstrating energy gains of $\sim \SI{1.5}{GeV}$ ($\sim \SI{2.5}{GeV}$) for drive pulse energies of \SI{2.4}{J} (\SI{5.0}{J}) are presented. Our results suggest that P-MoPAs driven by few-joule, picosecond pulses, such as those provided by high-repetition-rate thin-disk lasers, could accelerate electron bunches to multi-GeV energies at pulse repetition rates in the kilohertz range.

\end{abstract}

\maketitle

\section{Introduction}

First proposed in 1979 by Tajima and Dawson, laser wakefield accelerators (LWFA) \cite{PhysRevLett.43.267} can outperform the accelerating gradients of conventional radio-frequency cavities by up to three orders of magnitude. In a LWFA, an ultrashort laser pulse excites a plasma wave ``wake'' by pushing free electrons via the ponderomotive force whilst the heavier ions remain approximately stationary. The separation of electrons and ions generates large electric fields, which propagate with a phase velocity set by the group velocity of the laser pulse, and are therefore suitable for accelerating relativistic charged particles.

For a laser pulse to efficiently excite a plasma wave, its duration must be less than half the plasma period $T_p = 2 \pi / \omega_p$, which is in the 100 fs range for plasma densities of interest. Because sufficiently short joule-scale pulses were not yet feasible, Tajima and Dawson proposed the plasma beatwave accelerator (PBWA) \cite{PhysRevLett.43.267}. In this configuration, two long co-propagating pumps with a carrier frequency mismatch approximately equal to the plasma frequency $\Delta\omega=\omega_1-\omega_2\approx\omega_p$ interfere to form a train of short pulses via beatwave modulation. The resulting cosine-squared intensity modulation corresponds to a train of pulses spaced by $T_p$ that can resonantly excite a plasma wave \cite{PhysRevLett.43.267,PhysRevLett.29.701,10.1063/1.99996,doi:10.1063/1.859401}. More generally, we will refer to resonant wakefield excitation with trains of uniformly or non-uniformly spaced pulses as multipulse-LWFA (MP-LWFA) \cite{Hooker_2014}. 

The great majority of recent LWFA experiments have instead used single high intensity ultrashort $(<\SI{100}{fs})$ laser pulses from Ti:sapphire laser systems enabled by the development of chirped pulse amplification (CPA) \cite{STRICKLAND1985219}. At high pulse energies ($>\SI{1}{J}$), these laser systems are restricted to low ($\sim$ 0.1-10 Hz) repetition rates \cite{doi:10.1063/1.4773687} and have poor ($<0.1\%$) electrical-to-optical energy efficiencies \cite{PWASC}. These limitations reduce the number of applications for which these LWFAs offer an advantage over conventional, radio-frequency particle accelerators.

To tackle these issues, a new scheme was recently proposed  \cite{PhysRevLett.127.184801}, which we have dubbed the Plasma-Modulated Plasma Accelerator (P-MoPA). This approach aims to drive high repetition rate, multi-GeV plasma accelerators with thin-disk lasers which are both efficient and can provide multi-joule laser pulses at kHz repetition rates \cite{Herkommer:20,Nagel:21,Produit:21}. These laser systems cannot be used to drive LWFAs directly since the duration of the pulses they provide is too long: $\tau \gtrsim \SI{1}{ps}$ for joule-scale pulses \cite{Paschotta2001,Sudmeyer2009,Baer:10}. To overcome this limitation, in a P-MoPA the drive pulse is spectrally broadened by adding multiple sidebands spaced by the plasma frequency using a `modulator' stage. This is followed by a dispersive optical system that removes the spectral phase exhibited by the sidebands, to generate a train of short pulses that can be used to resonantly excite a wakefield in an accelerator stage.

In this paper we study the performance of multi-GeV MP-LWFAs driven by P-MoPA pulse trains; it builds on our earlier work that established the range of parameters for which operation of the modulator stage was stable \cite{PhysRevE.108.015204}. We derive a full 3D analytic theory of the pulse train and wakefield evolution in long plasma channels, and compare the results with PIC simulations using WarpX \cite{WarpX}. We use these results to optimize the performance of P-MoPAs, which are set by both the dynamics of the accelerator stage, as well as the previously derived constraints on the modulator stage \cite{PhysRevE.108.015204}. We explore the key differences between the pulse trains formed by P-MoPA and PBWA. We find that a beneficial feature of the P-MoPA scheme is the ability to control the longitudinal profile of the pulses in the train independent of their pulse spacing; this allows for more resilient pulse trains and more efficient wake excitation than PBWA, as well as offering a degree of control over the onset of depletion effects.

\begin{figure*}
    \includegraphics[width=\textwidth]{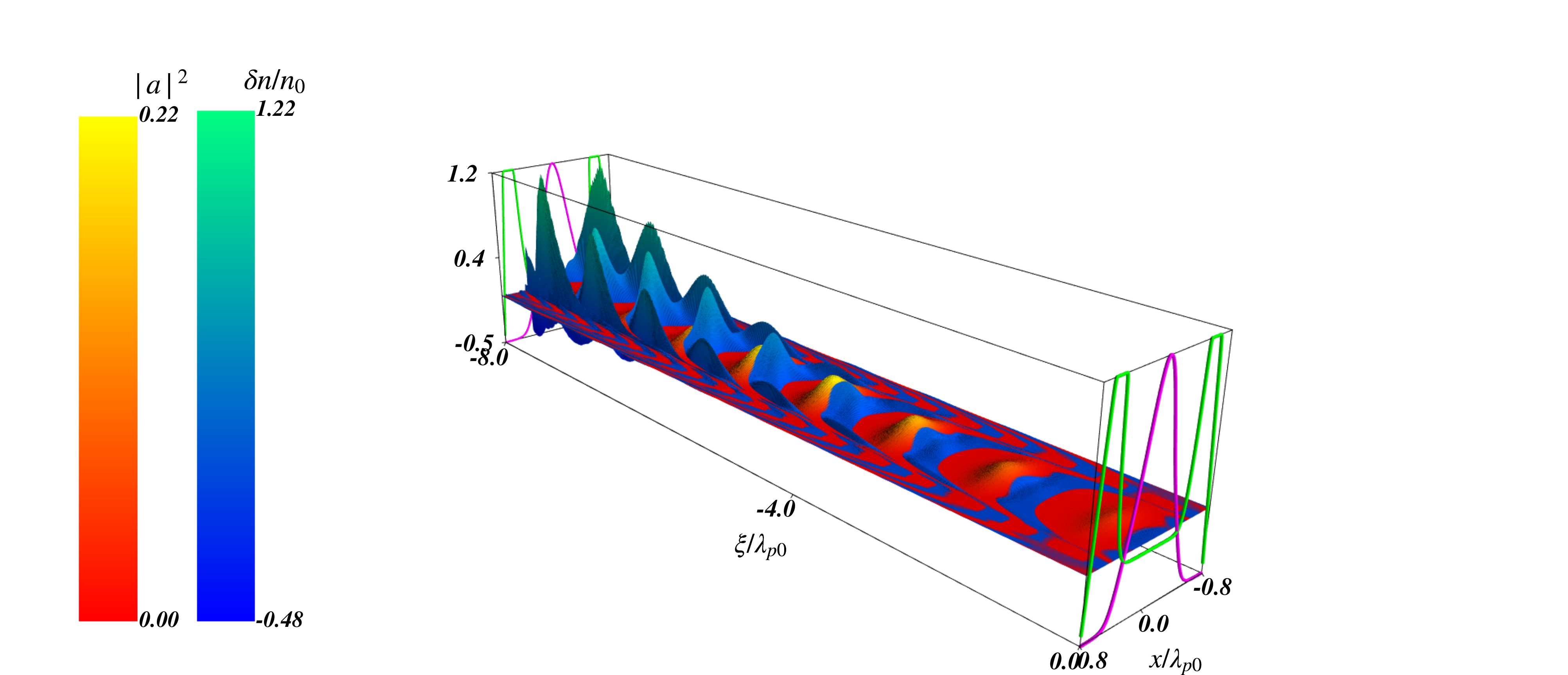}
    \caption{[Color online]. Snapshot at $z=\SI{2.1}{mm}$ of the relative plasma density perturbation $\delta n/n_0$ (blue-green) resonantly excited by a P-MoPA pulse train with normalized intensity $|a|^2$ (red-yellow). The transverse electron density profile of the channel, corresponding to Eq. (\ref{eq:quasisquare}) (green), and the longitudinally-averaged transverse intensity profile (purple) of the drive pulses are shown on the end panel of the plot. Results are from a 2D simulation using the PIC code WarpX \cite{WarpX} with $W_\text{drive}=\SI{1.2}{J}$, $\tau_\text{drive}=\SI{1}{ps}$, $\lambda_L=\SI{1030}{nm}$, $R=\SI{30}{\micro m}$, $w_0=\SI{30}{\micro m}$, $n_{00}=\SI{2.5e17}{cm^{-3}}$ and modulator parameter $\beta=1.2$ as defined in Eq. (\ref{eq:pmopa_envelope}).}
    \label{fig:3d}
\end{figure*}

\section{Scaling Laws}

Since the P-MoPA accelerator stage necessarily operates in the (quasi-)linear regime, we expect the energy gain scaling laws for a P-MoPA to be similar to those of the single pulse linear regime \cite{RevModPhys.81.1229}. In this section we outline these scaling laws, taking into account the effect of finite spot size effects on the laser group velocity.

To extend the acceleration length beyond the Rayleigh range, the pulse train is guided in a pre-formed plasma channel. A parabolic plasma channel matched to a Gaussian pulse of spot size $w_0$ has a transverse electron density profile given by \cite{1126872, FIRTH1977226, PhysRevE.59.1082}
\begin{align}
    n_0(r) = n_{00}+\Delta n(r/w_0)^2
\end{align}
where $n_{00}$ is the on-axis plasma density and $\Delta n = (\pi r_ew_0^2)^{-1}$ is the channel depth parameter with $r_e$ being the classical electron radius. The corresponding group velocity of a Gaussian pulse propagating in such a matched channel, including finite spot size effects, is given by \cite{PhysRevE.59.1082}
\begin{align}
    \frac{v_g}{c} = 1 - \frac{\omega_{p0}^2}{2\omega_L^2}-\frac{2c^2}{\omega_L^2w_0^2}\,,\label{Eqn:vg}
\end{align}
where $\omega_{p0} = \omega_p(r=0)$ indicates the on-axis plasma frequency. We note that the contributions to the reduction in $v_g$ of the finite spot size and plasma density are equal when $w_0 = 2 c / \omega_{p0}$, which for an on-axis plasma density of $n_{00}=\SI{1.0e17}{cm^{-3}}$ corresponds to $w_0=\SI{34}{\micro m}$.

From Eq. (\ref{Eqn:vg}), the dephasing length, defined as the propagation distance over which an electron propagating at $c$ slips relative to the wakefield by one-quarter plasma wavelength $\sim\lambda_{p0}/4$ (assuming no channel effects on wake structure), is
\begin{align}\label{eq:Ld}
     L_\text{d} = \frac{\lambda_{p0}}{4}\left(\frac{\omega_{p0}^2}{2\omega_L^2}+\frac{2c^2}{\omega_L^2w_0^2}\right)^{-1}.
\end{align}
Note that a method has been proposed \cite{PhysRevAccelBeams.23.021303} which allows MP-LWFAs to accelerate beyond the dephasing length by introducing a short region of higher plasma density, which lets the accelerated electron bunch to rephase into the next accelerating bucket back at the resonant density without deceleration.

Integrating the equations of motion for a relativistic particle moving in a sinusoidal wakefield gives the energy gain $\Delta W_e$ before the onset of dephasing as
\begin{align}\label{eq:W_e}
    \frac{\Delta W_e}{m_ec^2} \approx \frac{2}{\pi}\frac{eE_\text{max}L_\text{d}}{m_ec^2} = 2\,\frac{E_\text{max}}{E_\text{wb}}\frac{\omega_L^2}{\omega_{p0}^2+4c^2/w_0^2}
\end{align}
where $E_\text{max}$ is the peak accelerating field and $E_\text{wb}=m_e\omega_{p0}c/e$ is the cold wavebreaking field. This shows that for a given spot size, the maximum energy gain cannot be increased indefinitely by reducing the plasma density as the dephasing length becomes dominated by finite spot size effects. This is a well known effect in LWFA \cite{PhysRevSTAB.13.101301}, where the prescription is to scale the spot size with the plasma density such that
\begin{align}
    w_0 \gtrsim \frac{2c}{\omega_{p0}} = \frac{\lambda_{p0}}{\pi}\,.
\end{align}
On top of preventing an excessive reduction in group velocity, scaling the spot size this way also prevents the wakefield from becoming largely radial \cite{10.1063/1.1715427,PhysRevSTAB.13.101301}. As the required spot size increases with the plasma wavelength, the total laser energy must necessarily increase. The total energy of the pulse train $W_\text{drive}$ required to excite a given accelerating field relative to the wavebreaking field $E_\text{max}/E_\text{wb}$ scales with the volume enclosed by the spot size and the plasma length, i.e. $W_\text{drive}\sim w_0^2\lambda_{p0}\sim\lambda_{p0}^3$. This grows faster with the plasma wavelength than the electron energy gain $\Delta W_e\sim\lambda_{p0}^2$ and the maximum bunch charge $Q_e\sim n_{00}w_0^2\lambda_{p0}\sim \lambda_{p0}$, but the overall laser-to-bunch energy efficiency $Q_e\Delta W_e/eW_\text{drive}$ has no plasma density scaling. These scaling laws, which we have derived for pulse trains in the linear regime, are identical to that of a linear single pulse LWFA \cite{RevModPhys.81.1229}.

Another limit to acceleration is pump depletion. Similar to the dephasing length, the depletion length will also be reduced by finite spot size effects. This is due to the extra pulse energy spent exciting the aforementioned transverse wakefield for pulses with spot sizes comparable to the plasma wavelength. The depletion length can be estimated by comparing the laser energy with the energy contained within the linear wakefield and its transverse counterpart spanning the depletion length $L_\text{pd}$. Assuming that the pulse train is resonantly spaced and is comprised of individual linearly-polarized Gaussian pulses of the form $a_m^2=a_{0,m}^2\exp(-(\xi-\xi_{0,m})^2/L^2)$, where $m = 1, 2, \ldots N$, and that each pulse is short (i.e. $L \ll \lambda_{p0}$) so that each pulse increases the normalized wake amplitude \cite{Gorbunov} by $\sqrt{\pi/2}\,a_{0,m}^2 k_{p0}L/4$, the pump depletion length is given by
\begin{align}\label{eq:L_pd}
    L_\text{pd} = \left(\frac{\pi}{8}\frac{E_\text{max}}{E_\text{wb}}\right)^{-1}L_\text{d} \,.
\end{align}
Note that this derivation is identical to that for single pulse LWFA. This expression shows that the depletion length is always larger than the dephasing length in the linear wakefield regime, which is also true for single pulse LWFA \cite{RevModPhys.81.1229}. To make efficient use of the available laser energy, we would ideally want the pump depletion length to be larger but similar to the dephasing length. As this is not feasible, the P-MoPA accelerator stage should at least be operated in the quasi-linear regime, whilst avoiding nonlinear detuning effects associated with the large amplitude wake which will be discussed later. We also note that this treatment of the depletion length is too simplistic for pulse trains since, as we will show later, it does not capture the different rates at which each of the pulses deplete. 

\section{Channel Effects on Energy Gain}

Driving the accelerator within a pre-formed plasma channel causes the laser to remain focused well beyond the Rayleigh range. However, the transverse variation of plasma density within the channel leads to a concomitant variation of the plasma frequency that can also influence the structure of the excited wakefield and hence the energy gain of an injected electron bunch. 

It is well known for single-pulse LWFA that plasma channels can strongly influence wakefield structure via the curvature of wake phase-fronts \cite{ANDREEV1998469}. For example, shallow parabolic channels matched to the spot size $w_0$, i.e. $\frac{1}{2}k_{p0}^2w_0^2\gg 1$, increase the overlap between the accelerating and focusing regions of the wakefield by up to a factor of two at sufficient distances behind the laser pulse. This can be beneficial for acceleration as it can increase the dephasing length by a factor of up to two without sacrificing much accelerating gradient. However, deep matched parabolic channels, i.e. those for which $\frac{1}{2}k_{p0}^2w_0^2<1$, yield a rapid conversion of axial field into radial field with distance behind the drive pulse, meaning that LWFA is only possible in the first few buckets behind the drive pulse. This quick conversion of axial field into radial field readily outpaces other known wakefield decay mechanisms such as the ion motion modulational instability \cite{PhysRevE.107.L023201} which only becomes relevant on the ion plasma frequency timescale, corresponding to about forty plasma wavelengths for hydrogen. Therefore the transverse profile of the plasma channel is the dominant effect determining damping of the accelerating wakefield, and hence the useful number of pulses in the train. 

When using a single pulse driver, the channel effect on wake structure only needs to be considered for one plasma wavelength behind the driver. However, for a pulse train comprised of $N$ pulses, the channel influence on wake structure is relevant over $N$ plasma wavelengths. The depth of a parabolic channel will therefore determine the maximum useful number of pulses used in a MP-LWFA, as the axial wakefield contribution of the first pulse in a train will disappear by the time the $N^\text{th}$ pulse passes. There are two ways to allow for longer trains. The first way is to operate well in the shallow channel limit $\frac{1}{2}k_{p0}^2w_0^2\gg N$. However, this forces the required matched spot size to become large at low plasma densities and large pulse train lengths (e.g. $w_0\gg\SI{75}{\micro m}$ at $n_{00}=\SI{1.0e17}{cm^{-3}}, N=10$), bringing the required laser energy to drive the wakefield in the quasi-linear regime to the 10-joule scale for a $\SI{1030}{nm}$ laser wavelength. The other solution is to use square-profile channels, which have a much weaker effect on the wakefield structure. Fortunately, the transverse density profiles of CHOFI plasma waveguides are closer to square than parabolic in shape \cite{PhysRevE.102.053201,Feder.2020,Miao.2020}, which makes them well suited to driving P-MoPAs. 

An example PIC simulation of the wake structure driven by a P-MoPA pulse train is presented in Figure \ref{fig:3d}, where the pulse train is guided in a quasi-square channel of the form
\begin{align}\label{eq:quasisquare}
    &\frac{n_0(r)-n_{00}}{\Delta n}= \nonumber \\
    &\begin{cases}
        (r/R)^{10} & r < 1.2R \\
        (1.2)^{10} & 1.2R\leq r < 1.2R+d \\
        (1.2)^{10}\left(1-\frac{r-1.2R-d}{d}\right) & 1.2R+d\leq r < 1.2R+2d \\
        0 & \text{otherwise}
    \end{cases}
\end{align}
where $R$, $\Delta n = (\pi r_e R^2)^{-1}$ and $d=\SI{10}{\micro m}$ are the channel wall radius, depth and thickness parameters respectively. Resonant excitation of a large amplitude $\sim 50\%$ wakefield on axis is evident. Also apparent is a large amplitude wakefield on the inner wall of the channel. This is due to the steep plasma density gradient
\begin{align}
    \frac{\mathrm{d}\ln n_0}{\mathrm{d}r} > k_p\,,
\end{align}
which is characteristic of plasma channels at these spot sizes and low $\sim10^{17}$ cm$^{-3}$ on-axis densities. The wake amplitude in this region continues to grow even after the laser driver has already passed due to the aforementioned conversion of the wakefield from the longitudinal to transverse component \cite{ANDREEV1998469}.
This accumulated wall-boundary wake can eventually exceed the on-axis wake amplitude, suggesting that large-amplitude wakefields driven by pulse trains in plasma channels would undergo transverse wavebreaking effects before longitudinal wavebreaking.

\section{P-MoPA Pulse Shaping}

The plasma modulator offers a degree of control of the temporal profile of the pulses in the pulse train, allowing for further optimization of the accelerator stage. The pulse train formed after compression will be comprised of pulses evenly spaced by the modulator plasma period with pulse heights and durations determined by the amount of spectral modulation introduced by the modulator stage. The additional bandwidth of the spectrally-modulated drive pulse enables a P-MoPA to form pulses substantially shorter than their separation, unlike the cosine-squared pulses created with the beatwave method. This allows larger wake amplitudes to be driven in a P-MoPA with the same drive pulse energy.

\begin{figure}
    \centering
    \subfloat[\label{fig:2a}]{
        \includegraphics[width=1.0\linewidth]{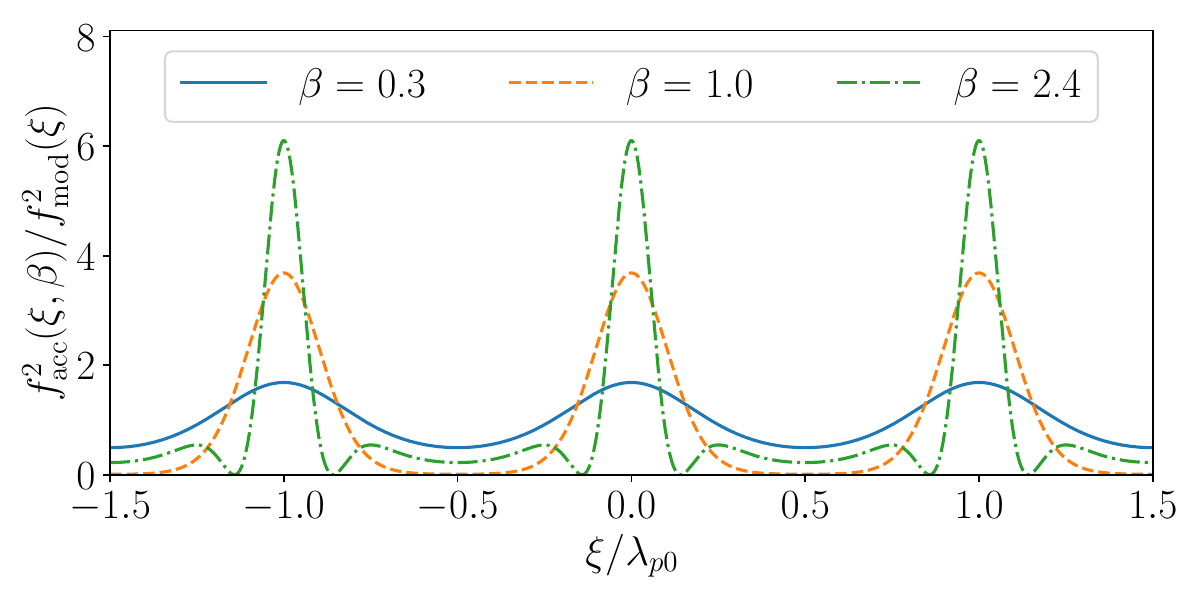}
    }
    \hspace{1cm}
    \subfloat[\label{fig:2b}]{
        \includegraphics[width=1.0\linewidth]{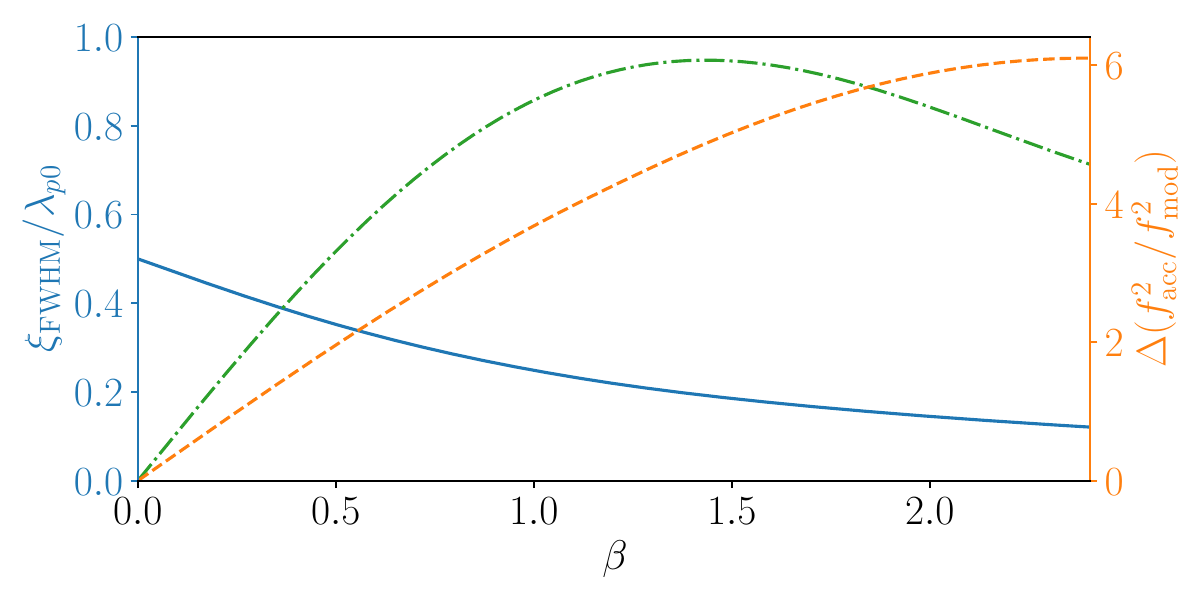}
    }
    \caption{[Color online]. (a) Pulse train temporal profiles for various values of the modulator parameter $\beta$ and their respective FWHM durations (dashed). (b) FWHM duration (blue, solid), peak prominence (orange, dashed) and the wakefield amplitude according to 1D linear theory (green, dot-dashed, arb. units) as a function of $\beta$.}
    \label{fig:pmopa_pulses}
\end{figure}

According to 3D spectral modulation theory \cite{PhysRevE.108.015204}, assuming that the spectral modulation remains primarily in the fundamental channel mode, the pulse trains formed using the ``ideal compressor'' described by Jakobsson et al \cite{PhysRevLett.127.184801}, with a seed wake of the form $\delta n(r,\xi) = \delta n_s\cos(k_{p0}\xi)\exp(-2r^2/w_0^2)$, will have the following temporal amplitude envelope
\begin{align}\label{eq:pmopa_envelope}
    &f_\text{acc}(\xi;\beta) =
    f_\text{mod}(\xi)\left(J_0(\beta)+2\sum_{n=1}^{\infty}J_n(\beta)\cos(nk_{p0}\xi)\right)\,, \nonumber \\
    &\beta = \frac{2\Omega_sL_\text{mod}}{v_{g,\text{mod}}}
\end{align}
where $\xi=z-ct$ is the longitudinal co-moving coordinate, $0\leq f_\text{mod}(\xi)\leq1$ is the input temporal amplitude envelope, $L_\text{mod}$ is the modulator length,  $v_{g,\text{mod}}$ is laser group velocity, and $\Omega_s=(\omega_{p0}^2/8\omega_L)(\delta n_s/n_{00})$ is the rate of spectral modulation of a Gaussian pulse co-propagating with a seed wake of on-axis amplitude $\delta n_s$ in a parabolic channel \cite{PhysRevE.108.015204}. The modulator parameter $\beta$ determines the effective number of sidebands generated in the drive pulse spectrum and controls the temporal profile of the pulse train produced by the plasma modulator (whilst obeying the modulator stability condition outlined in \cite{PhysRevE.108.015204}). Figure \ref{fig:pmopa_pulses} shows how the temporal intensity profiles of the pulses within the train, their full-length at half maximum duration, and their contrast, varies with $\beta$. It can be seen that adjusting the parameter $\beta$ allows the duration and contrast of the pulses within the train to be controlled.

The peak intensity occurs at $\beta=j_{0,1}\approx2.405$, where $j_{m,n}$ indicates the $n^{\text{th}}$ nonzero root of the Bessel function $J_m(x)$. This can be shown by evaluating the temporal envelope at the pulse center using well-known identities for Neumann series of Bessel functions \cite{watson1966,gradshteyn2007}
\begin{align}
    \frac{f_\text{acc}(0;\beta)}{f_\text{mod}(0)} &=
    J_0(\beta)+2\sum_{n=1}^{\infty}J_n(\beta) \nonumber \\
    &= 1 + \int_0^\beta J_0(s)ds
\end{align}
which shows that its global maximum occurs at $\beta=j_{0,1}$, where it takes the value $f_\text{acc} \approx 2.470\,f_\text{mod}$. As evident in Figure \ref{fig:pmopa_pulses}, increasing $\beta$ above this value leads to the formation of shorter pulses, but these have temporal wings, which wastes energy since they do not contribute to wake excitation. Thus when optimizing the P-MoPA scheme we only need to consider $\beta\leq j_{0,1}$.

The additional control that the parameter $\beta$ provides enables more efficient wake excitation than is possible with PBWA. This is shown in Figure \ref{fig:pmopa_pulses}(b), which shows that $\beta$ in the range of 1.1--1.6 optimizes the wake amplitude according to linear wakefield theory \cite{RevModPhys.81.1229}. We can also show in the linear regime that the ratio between the amplitude of the wake potential driven by a P-MoPA pulse train and an ideal beatwave pulse train with a temporal intensity modulation of the form $1+\cos(k_{p0}\xi)$ containing the same laser energy is given by
\begin{align}
    \frac{\phi_\text{P-MoPA}}{\phi_\text{PBWA}} &= \frac{\int_{-\pi}^\pi \left[J_0(\beta)+2\sum_{n=1}^{\infty}J_n(\beta)\cos(ns)\right]^2\cos(s)ds}{\int_{-\pi}^\pi \left[1+\cos(s)\right]\cos(s)ds}\nonumber \\
    &= 4\sum_{n=0}^\infty J_n(\beta)J_{n+1}(\beta) \nonumber \\
    &= 2\beta\left[J_0^2(\beta)+J_1^2(\beta)\right]\,.
\end{align}
This ratio is maximized for $\beta\approx 1.43$, at which value the wake amplitude in a P-MoPA is $72\%$ larger than that driven by an ideal PBWA with the same total drive energy.

\captionsetup[subfloat]{position=top,justification=raggedright,singlelinecheck=false}
\captionsetup[subfloat]{position=bottom,justification=centering,singlelinecheck=true}

\section{The Rosenbluth-Liu Limit}

Resonant wakefield acceleration requires the driving pulses to be spaced by the plasma period. However, as the plasma wave amplitude increases beyond the linear regime, its wavelength increases due to relativistic effects. The consequent build-up of plasma wave phase difference relative to the laser pulse centroids sets a limit on the maximum wake amplitude achievable. 

The saturation effect arising from this detuning was first derived by Rosenbluth and Liu \cite{PhysRevLett.29.701} in the context of plasma beatwave excitation. In their derivation, two infinite-duration pumps with normalized vector potential amplitudes $a_1$ and $a_2$ interfere to form a cosine beatwave pulse train resonant with the plasma, which results in the saturated wakefield amplitude scaling with the third root of the intensity of the beatwave modulation
\begin{align}
    \frac{E_\text{sat}}{E_\text{wb}} = \left(\frac{16a_1a_2}{3}\right)^{1/3}.
\end{align} 
P-MoPA pulse trains, which are also spaced uniformly by design, are thus expected to exhibit a similar limit. However, for a P-MoPA, an important difference is that the train will comprise a relatively small number of pulses. 

According to 1D nonlinear plasma wave theory \cite{RevModPhys.81.1229}, the nonlinear correction to the plasma wavelength for a plasma wave of amplitude $\phi_a=E_a/E_\text{wb}$ at non-relativistic laser intensities in the $\phi_a^2\ll1$ limit is given by
\begin{align}\label{eq:kp_NL}
    \lambda_p^\text{NL} \approx \lambda_p(1+3\phi_a^2/16)\,, \nonumber \\
    k_p^\text{NL} \approx k_p(1-3\phi_a^2/16)\,.
\end{align}
Resonant excitation of the wakefield eventually saturates due to detuning caused by the progressive growth of the plasma wavelength. The co-moving coordinate $\xi_\text{sat}$ at which this detuning occurs is approximately given by
\begin{align}\label{eq:detuning}
    \int_{\xi_\text{sat}}^\infty(k_p-k_{p}^\text{NL})d\xi \approx \frac{\pi}{2}\,.
\end{align}
Since the pulse train in a P-MoPA pulse train corresponds to a sequence of short pulses spaced by $T_p$, within an envelope determined by the duration of the ps-duration drive pulse, most of its energy will be contained within only a few pulses close to its center. To account for this we model the original pulse envelope as a temporally Gaussian pulse of the form
\begin{align}
    f_\text{mod}^2(\xi) = \exp\left(-\frac{\xi^2}{2\sigma^2}\right)\,,
\end{align}
where $\sigma = c\tau_\text{drive}/2\sqrt{2\ln2}$ with $\tau_\text{drive}$ being the FWHM duration of the drive pulse. From linear theory, ignoring detuning, the wakefield potential amplitude $\phi_a(\xi)$ excited by the pulse train is given by
\begin{align}\label{eq:phi_s}
    \frac{\phi_a(\xi)}{\phi_\text{max}} = \frac{\int_\xi^\infty d\xi'f_\text{acc}^2(\xi')}{\int_{-\infty}^\infty d\xi'f_\text{acc}^2(\xi')} \approx \frac{1}{2}\,\text{erfc}\left(\frac{\xi}{\sqrt{2}\,\sigma}\right)\,,
\end{align}
where $\phi_\text{max}=E_\text{max}/E_\text{wb}$ is the maximum wakefield amplitude resonantly excited by the pulse train. For the pulse train to efficiently drive a wakefield, resonant detuning must not occur before the majority of the pulses have already passed. If we demand that at least $\sim90\%$ of the pulse train energy passes before the full $\pi/2$ detuning occurs, then combining Eqs. \ref{eq:kp_NL}, \ref{eq:detuning} and \ref{eq:phi_s} results in the following condition
\begin{align}
    -\frac{\xi_\text{sat}^{90\%}}{\lambda_p} > 0.27\,N_\text{eff}\,,
\end{align}
where $N_\text{eff} = 2\tau_\text{FWHM}/T_p$ is defined as the effective number of pulses in the pulse train. Substituting this expression along with Eq. (\ref{eq:phi_s}) into Eq. (\ref{eq:detuning}) yields the resonant detuning limit for wakefields driven by P-MoPA pulse trains:
\begin{align}\label{eq:phi_max}
    \phi_\text{max}^{90\%} = \frac{E^{90\%}_\text{max}}{E_\text{wb}} < \frac{2.8}{\sqrt{N_\text{eff}}}\,.
\end{align}
For example, for $\tau_\text{FWHM}=\SI{2}{ps}$ at $n_{00}=\SI{2.5e17}{cm^{-3}}$, a P-MoPA would form a pulse train comprised of $N_\text{eff}=18$ pulses. Substituting this into Eq. (\ref{eq:phi_max}) then limits the maximum accelerating gradient before detuning to $E_\text{max}<\SI{32}{GV/m}$, or $< 66\%$ of the cold wavebreaking field. If we operate the P-MoPA in the linear and quasi-linear regime, for which $\phi^{90\%}_\text{max}\lesssim0.5$, then Eq. (\ref{eq:phi_max}) becomes $N_\text{eff}<30$. We therefore conclude that wakefield saturation caused by this detuning mechanism will not be an issue for a drive train comprising $\sim 30$ or fewer pulses.

A further consideration for P-MoPA pulse trains is that the individual pulses can be much shorter than in PBWA, such that $a\ll1$ can no longer be assumed. Thus, to maximize the wake amplitude, the relativistic effects associated with the quiver motion in regions of high intensity, rather than the relativistic effects associated with the fluid motion, should also be corrected for by using a slightly higher plasma density in the accelerator than the modulator stage to counteract this detuning mechanism.    

Aside from limiting the wakefield amplitude, resonant detuning can also lead to focusing/defocusing effects caused by pulses co-propagating with high amplitude regions of the wakefield. These effects are discussed in the next section.

\section{Relativistic Self-Phase Modulation and Transverse Mode Excitation}

Up until now we have been treating the pulse train envelope as unchanging as it propagates along the accelerator. While this assumption is useful for deriving scaling laws and isolating certain physics, in reality the pulse train is a dynamic 3D structure which will evolve under the influence of both relativistic nonlinearities and its own excited wake. Neglecting depletion for now, the two main effects to consider for MP-LWFA in long channels are relativistic self-phase modulation (SPM) and tranvserse mode dynamics caused by the excitation of higher order transverse channel modes by the focusing/defocusing effects of the large amplitude wake. 

Relativistic SPM causes further red/blue-shifting near the head/tail of each pulse, especially for the shortest duration, highest $a_0$ pulses. Transverse mode excitation results in spot size oscillations and hence oscillations in the wakefield structure, which become more pronounced towards the back of the pulse train where the wake amplitude is the greatest. These oscillations could negatively affect beam quality, since a propagation-varying focusing field could increase the energy spread and emittance of the accelerated bunch.

Before significant pump depletion has occurred, i.e. while the frequencies within the laser spectrum satisfy $|\omega-\omega_L|/\omega_L\ll1$, the evolution of the pulse train propagating in an axisymmetric channel of the form $n_0(r)=n_{00}+\delta n_0(r)$ can be described using the paraxial wave equation in the weakly relativistic limit \cite{doi:10.1063/1.860707,NEAndreev_1994,doi:10.1063/1.872134,doi:10.1063/1.3691837}
\begin{align}\label{eq:GNLS}
    &\left[\frac{i}{\omega_L}\frac{\partial}{\partial\tau} + \frac{c^2}{2\omega_L^2}\Delta_\perp\right] a = \nonumber \\ &\frac{\omega_p^2}{2\omega_L^2n_0}\left[\delta n_0(r) + \delta n(r,\xi;|a|^2) - n_0(r)|a|^2/4\right]a
\end{align}
where $a(r,\theta,\xi,\tau)$ is the envelope of the normalized vector potential of the pulse, $\omega_L$ is the laser frequency and the propagation is described in co-moving coordinates $\xi=z-v_{g0}t$, $\tau=t$, with $v_{g0}/c = (1-\omega_{p0}^2/\omega_L^2)^{1/2}$ defined as the group velocity of electromagnetic plane waves in uniform plasma of density $n_{00}$, corresponding to the on-axis plasma channel frequency $\omega_{p0}$, and $\Delta_\perp = \partial^2/\partial r^2 + (1/r)\partial/\partial r + (1/r^2)\partial^2/\partial\theta^2$ is the tranvserse Laplacian.

\begin{figure*}[tb]
    \centering
    \includegraphics[width=0.49\textwidth]{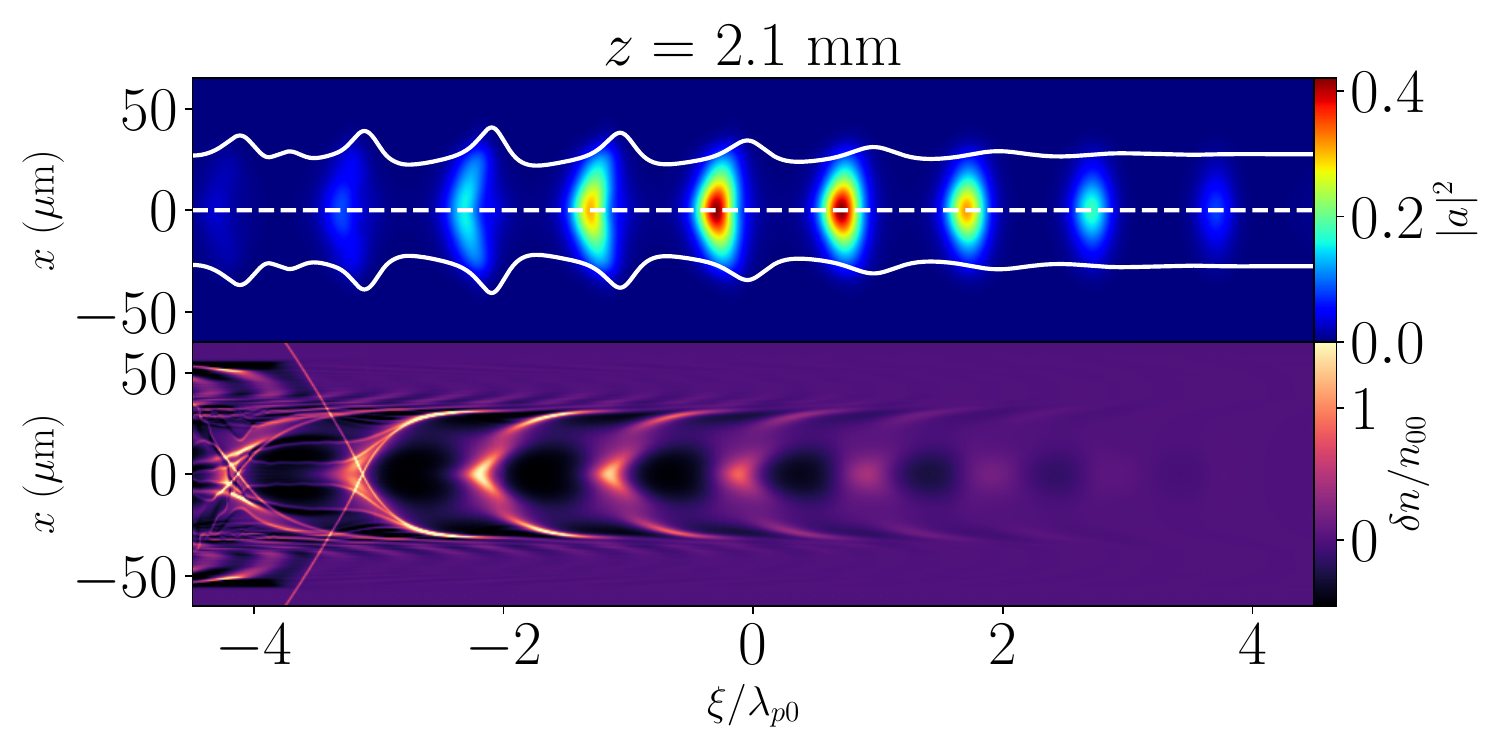}
    \includegraphics[width=0.49\textwidth]{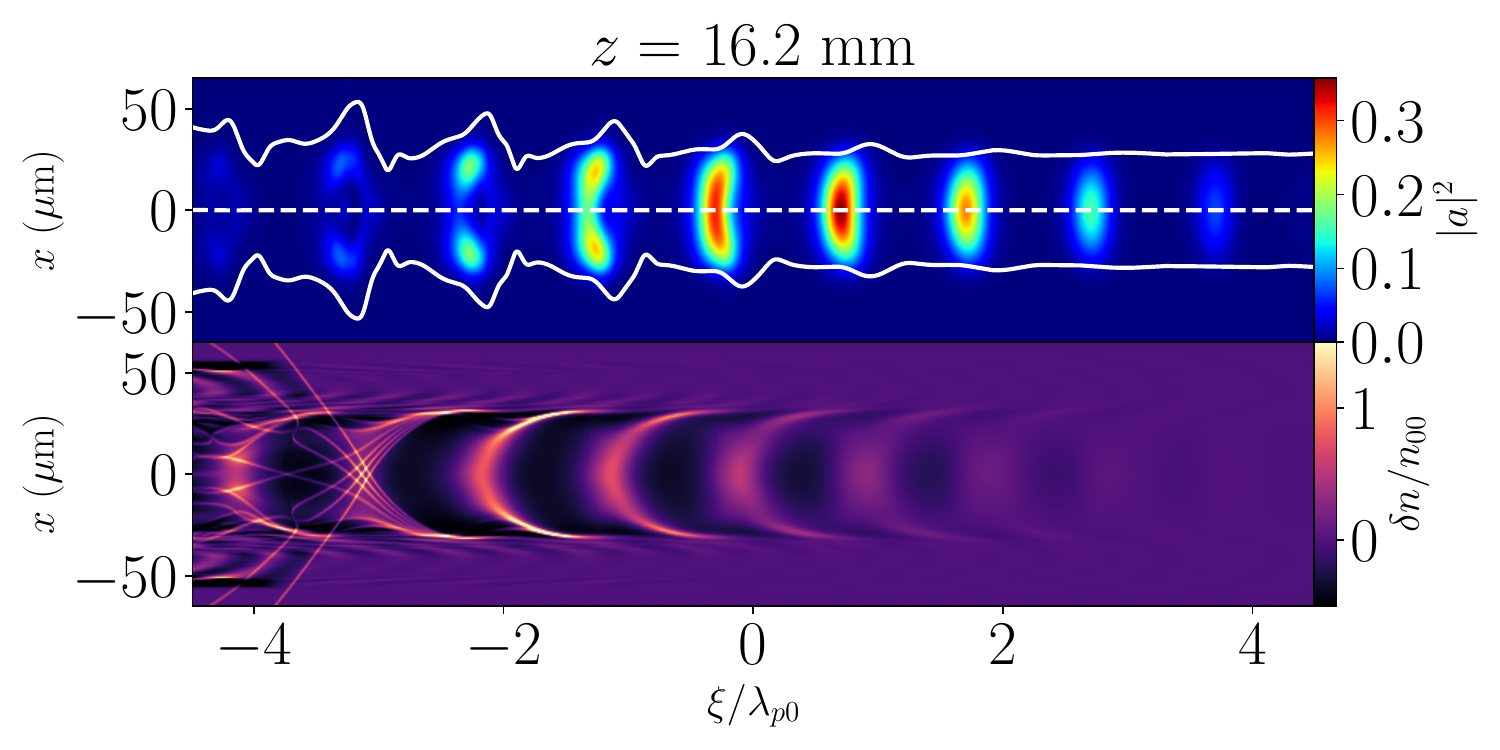}
    \includegraphics[width=0.49\textwidth]{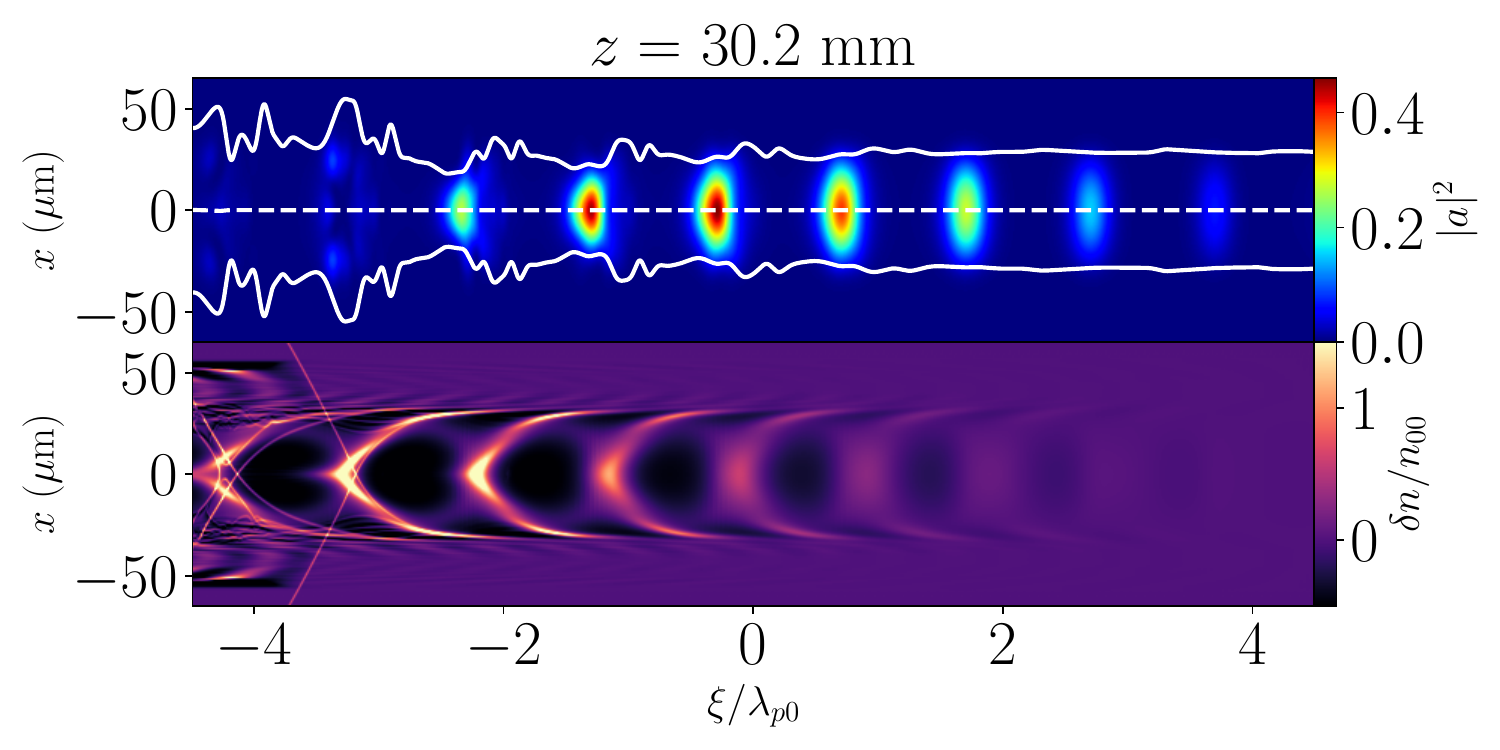}
    \includegraphics[width=0.49\textwidth]{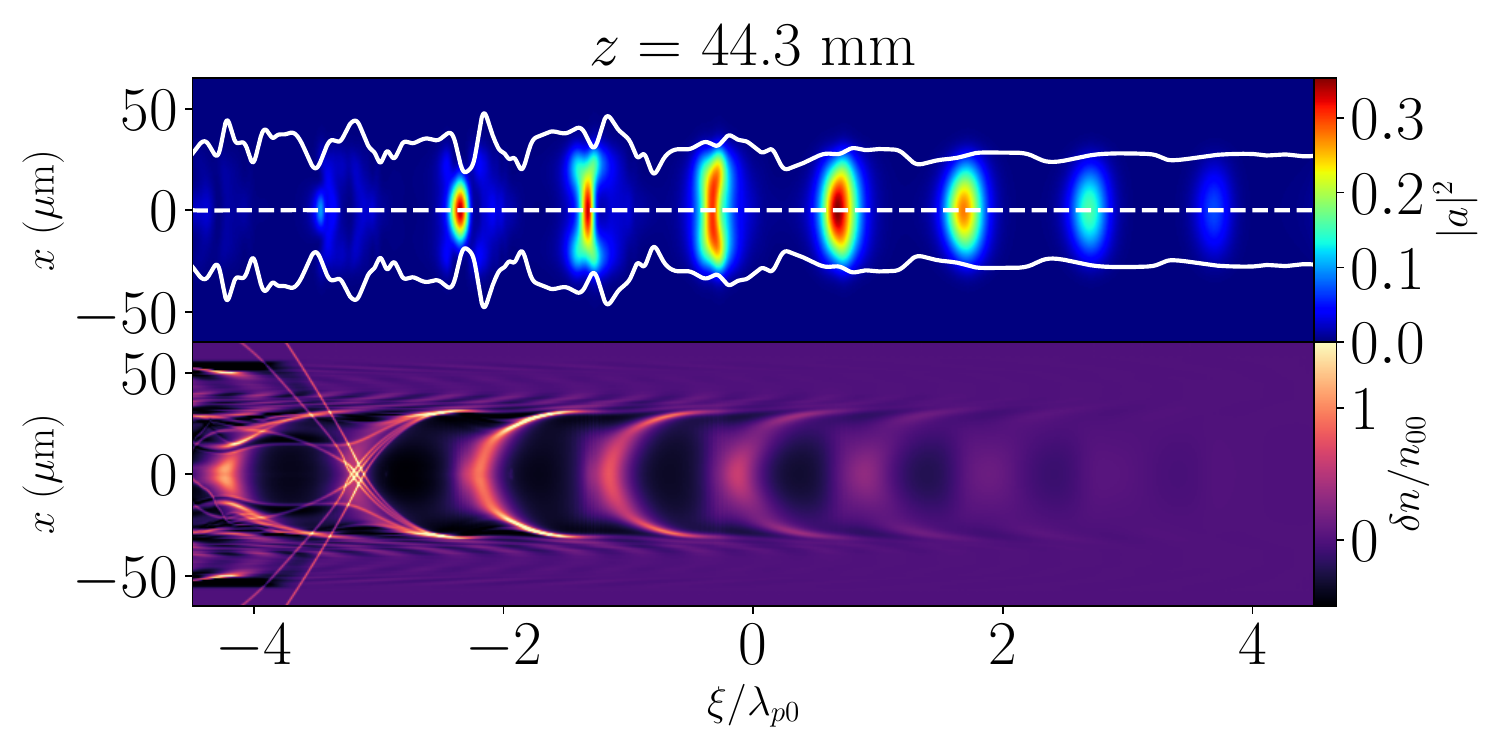}
    \caption{[Color online]. Results of a 2D PIC simulation of the propagation of a pulse train in a quasi-square plasma channel. For this simulation, $\xi$ as defined after Eq. (\ref{eq:GNLS}) at varying propagation distances $z$ in a quasi-square channel of the form given in Eq. (\ref{eq:quasisquare}) with $n_{00}=\SI{2.5e17}{cm^{-3}}$, $\Delta n=\SI{1.26e17}{cm^{-3}}$, $R=\SI{30}{\micro m}$, $W_\text{drive}=\SI{2.4}{J}$, $\tau_\text{drive}=\SI{1}{ps}$ and modulator parameter $\beta=1.2$. The pulse train at the entrance of the channel initially has a Gaussian transverse profile with spot size $w_0=\SI{30}{\micro m}$. For each plot, the top panel displays the laser intensity profile $|a|^2$ along with the local longitudinal variation of the laser centroid $x_c(\xi)$ (dashed) and local effective spot size $w_\text{eff}(\xi)$ (solid). The bottom panel displays the wake density perturbation $\delta n/n_{00}$.}
    \label{fig:TME}
\end{figure*}
Pre-formed plasma channels have often been modelled using parabolic transverse profiles, since these guide Laguerre-Gaussian modes. The long ($\gtrsim\SI{100}{mm}$), low-attenuation and kHz production rate plasma channels required for this scheme have been experimentally demonstrated using the CHOFI technique \cite{PhysRevE.102.053201,Feder.2020,Miao.2020}, which forms deep, low-loss channels with on-axis densities as low as $\SI{1e17}{cm^{-3}}$. These CHOFI channels are often better modelled with transverse profiles closer to square rather than parabolic \cite{Feder.2020}, meaning that wake phase-front curvature effects can be neglected. In the following analysis we will use a simplified profile and treat the plasma channel as an infinite square well of the form
\begin{align}\label{eq:squareCh}
    n_0(r) = 
    \begin{cases}
        n_{00} & r\leq R_\infty \\
        \infty & r>R_\infty
    \end{cases}\,,
\end{align}
where $R_\infty$ indicates the wall radius parameter specifically for this infinite square well. Note that the analysis that follows can be generalized to any axisymmetric, monotonically increasing channel profile. Assuming that the channel is unperturbed and that relativistic effects are negligible, substituting Eq. (\ref{eq:squareCh}) into Eq. (\ref{eq:GNLS}) yields the following orthogonal set of Bessel mode solutions for $r\leq R_\infty$
\begin{align}
    &a_{pm}(r,\theta,\xi,\tau) = \alpha_{pm}(\xi)\mathcal{J}_{pm}(r,\theta)e^{-i\Omega_{pm}\tau}\,, \nonumber \\
    &\mathcal{J}_{pm}(r,\theta) = \frac{J_1(j_{0,1})}{\left|J_{|m|+1}(j_{|m|,p+1})\right|}J_{|m|}\left(j_{|m|,p+1}\frac{r}{R_\infty}\right)e^{im\theta}\,, \nonumber \\
    &\Omega_{pm} = \frac{j_{|m|,p+1}^2c^2}{2\omega_LR_\infty^2}\,,
\end{align}
where the integers $p\geq0$ and $m$ are the radial and azimuthal indexes respectively. Note that the effective spot size $w_\text{0,eff}$ of the fundamental mode is related to the wall radius $R_\infty$ by
\begin{align}
    w_\text{0,eff} = \sqrt{\frac{\int_0^{R}2r^2\mathcal{J}_{00}^2(r)rdr}{\int_0^{R}\mathcal{J}_{00}^2(r)rdr}} \approx 0.66R_\infty\,.
\end{align}
which is analogous to the result for the standard step-index waveguide in the limit of infinite $V$-number \cite{Brinkmeyer:79}. Similar to a previous analysis of the P-MoPA modulator stage \cite{PhysRevE.108.015204}, long-propagation of pulse trains in plasma channels are expected to excite higher order transverse channel modes, with an amplitude that scales with the wake amplitude. To model this effect, we first assume a fixed multipulse-driven wake of the form
\begin{align}
    \delta n(r,\xi) = \delta n_a(\xi)\mathcal{J}_{00}^2(r) \,,
\end{align}
which is consistent with a pulse train primarily in the fundamental mode exciting a wake in the (quasi-)linear regime. Applying time-dependent perturbation theory (TDPT) to Eq. (\ref{eq:GNLS}) for pulses propagating in a square channel with only the $m=0$ radial transverse modes present, we find the following set of PDEs for the coefficients $\alpha_p(\xi,\tau)$ of each of the square channel's radial modes $\mathcal{J}_{p0}(r)$
\begin{align}\label{eq:TDPT} 
    &i\frac{\partial \alpha_p(\xi,\tau)}{\partial\tau} = \nonumber \\
    &\frac{\omega_{p0}^2}{2\omega_L}\sum_n \alpha_n(\xi,\tau)\Big\langle\mathcal{J}_{p0}\Big|\frac{\delta n}{n_{00}}-\frac{|a|^2}{4}\Big|\mathcal{J}_{n0}\Big\rangle e^{i\left(\Omega_{p0}-\Omega_{n0}\right)\tau} \,, \nonumber \\
    &a(r,\xi,\tau) = \sum_p \alpha_p(\xi,\tau)\mathcal{J}_{p0}(r)e^{-i\Omega_{p0}\tau}\,,\nonumber \\
    &\langle\mathcal{J}_{p'm'}|(...)|\mathcal{J}_{pm}\rangle \equiv \frac{\int_0^{2\pi}d\theta\int_0^{R}rdr\mathcal{J}_{p'm'}^\ast (...)\mathcal{J}_{pm}}{J_1^2(j_{0,1})\pi R_\infty^2}\,.
\end{align}
We note that TDPT requires that $\pi r_e(2R_\infty/j_{0,1})^2\delta n\ll 1$, which is \emph{not} true everywhere in the accelerator stage due to the large amplitude wake. However, for a resonant pulse train the zeros of $\delta n_a(\xi)$ are located near the laser centroids, so TDPT \emph{is} still valid provided that the duration of each pulse is short, so that $\delta n$ is sufficiently small in regions of high intensity. Hence we can still use TDPT to describe the evolution of the pulse train despite the large amplitude wake.

Assuming that most of the light is initially in the fundamental mode, transverse mode transitions will be dominated by excitations of only the first radial mode, with higher order mode excitations being much weaker. Hence substituting our expression for the wake and approximating Eq. (\ref{eq:TDPT}) as a two-level system description of the fundamental and first radial transverse modes yields 
\begin{align}\label{eq:spotosc}
    &\frac{\partial}{\partial\tau}
    \begin{pmatrix}
    \alpha_0 \\
    \alpha_1
    \end{pmatrix}
    = \bm{D}(\xi,\tau) 
    \begin{pmatrix}
    \alpha_0 \\
    \alpha_1
    \end{pmatrix}\,,\nonumber \\
    &\bm{D}(\xi,\tau) = -i\gamma(\xi)
    \begin{pmatrix}
    d_{00} & d_{01}e^{-i\omega_w\tau} \\
    c.c. & d_{11}
    \end{pmatrix}\,,  \nonumber \\
    &\gamma(\xi) = \frac{\omega_{p0}^2}{2\omega_L}\left(\frac{\delta n_a(\xi)}{n_{00}}-\frac{|a_0|^2f_\text{acc}^2(\xi)}{4}\right)\,,\nonumber \\
    &\omega_w = \Omega_{10}-\Omega_{00}\approx \frac{12.3\,c^2}{\omega_LR_\infty^2}\,\nonumber \\
    &d_{00} = \langle\mathcal{J}_{00}|\mathcal{J}_{00}^2|\mathcal{J}_{00}\rangle \approx 0.57\,,\,\,\, d_{01} = \langle\mathcal{J}_{00}|\mathcal{J}_{00}^2|\mathcal{J}_{10}\rangle \approx 0.27 \,,\nonumber \\
    &d_{11} = \langle\mathcal{J}_{10}|\mathcal{J}_{00}^2|\mathcal{J}_{10}\rangle \approx 0.48
\end{align}
where $\omega_w$ is the spot size oscillation frequency, $c.c.$ indicates the complex conjugate of the upper anti-diagonal entry, and we have assumed the fundamental intensity profile $|a|^2\approx |a_0|^2f_\text{acc}^2(\xi)\mathcal{J}^2_{00}(r)$ for the weakly relativistic term. This linear system of PDEs can then be solved analytically
\begin{align}\label{eq:analytic}
    &\begin{pmatrix}
    \alpha_0(\xi,\tau) \\
    \alpha_1(\xi,\tau)
    \end{pmatrix}   
    =
    \bm{A}(\xi,\tau)
    \begin{pmatrix}
    \alpha_0(\xi,0) \\
    \alpha_1(\xi,0)
    \end{pmatrix}\,, \nonumber \\
    &\bm{A}(\xi,\tau) = 
    \exp\left[-i\gamma(\xi)\tau
    \begin{pmatrix}
    d_{00} & d_{01}\text{sinc}\left(\frac{\omega_w\tau}{2}\right)e^{-i\omega_w\tau/2} \\
    c.c. & d_{11}
    \end{pmatrix}
    \right]\,,\nonumber \\
    &w_\text{eff}(\xi,\tau) \approx w_{0,\text{eff}}\left(1 - 0.40\,\frac{\alpha_0^\ast \alpha_1 e^{-i\omega_w\tau}+c.c.}{|\alpha_0|^2}\right) \,.
\end{align}
This solution describes the excitation of the first radial transverse mode due to transverse (de-)focusing and self-modulation, both of which can arise through the effects of the wake and self-focusing. The interference between the fundamental and first radial mode then leads to a modulation of the local effective spot size $w_\text{eff}(\xi,\tau)$. Evaluating this matrix exponential yields a unitary matrix with diagonal entries of $\mathcal{O}(1)$ and anti-diagonal entries of $\mathcal{O}(\gamma/\omega_w)\ll1$. This ensures that, as long as the original TDPT condition is satisfied in regions of high intensity, laser light initially in the fundamental transverse channel mode will also remain primarily in the fundamental mode, with small $\mathcal{O}(\gamma^2/\omega_w^2)$ and $\mathcal{O}(|\delta w|^2/w_0^2)$ oscillating exchanges in energy between the fundamental and first radial modes, where $\delta w/w_0\ll1$ is some small initial spot size mismatch.

Since the bulk of the wake is already excited by the zeroth order intensity $\sim\mathcal{J}_{00}^2(r)$, we can neglect the wake contribution of the $\mathcal{O}(\gamma/\omega_w)\ll1$, $\sim\mathcal{J}_{00}(r)\mathcal{J}_{10}(r)$ first order correction to the laser transverse mode dynamics. However, the near-axis wakefield itself can be significantly affected as laser energy continuously oscillates towards and away from the axis, resulting in propagation-dependent accelerating and focusing fields which could, in principle, impact both the emittance and energy spread of an injected electron bunch. 

An example of this phenomenon in 2D PIC simulations is evident in Figure \ref{fig:TME}, which shows snapshots of the laser intensity envelope and density perturbation as it propagates in a quasi-square channel. We can see that transverse mode excitation is the strongest for the trailing pulses which co-propagate with the largest amplitude wake, as described by Eq. (\ref{eq:analytic}). This results in an oscillation of the wake structure and hence the maximum on-axis accelerating field $E_\text{max}$. (This oscillation of on-axis field can also be observed directly in Figure \ref{fig:spectra}, as we discuss below.) Despite the transverse mode excitation causing a break-up of the trailing pulses, Figures \ref{fig:TME} and \ref{fig:centroids} show that the pulse train is still able to resonantly excite a wakefield even in the quasi-linear regime with $E_\text{max}/E_\text{wb}\sim0.5$.

To compare the spot size oscillation frequency $\omega_w$ for a square channel given in Eq. (\ref{eq:spotosc}) with the PIC simulations presented in Figure \ref{fig:spectra}, a correction has to be made for the fact that the plasma channel used in all of the presented PIC simulations is quasi-square with walls of finite height and width of the form given in Eq. (\ref{eq:quasisquare}). For this quasi-square channel we define an effective channel radius, $R_\infty^\text{eff}$, as that radius for which the fundamental mode, $\mathcal{J}_{00}(r)$, of an infinite square channel best matches the lowest-order mode of the quasi-square channel. For the $R=\SI{30}{\micro m}$ case, this results in an effective channel radius of approximately $R_\infty^\text{eff}\approx\SI{46}{\micro m}$. Substituting this into Eq. (\ref{eq:spotosc}) then gives a spot size oscillation wavelength of $2\pi c/\omega_w\approx\SI{6.6}{mm}$, which is in good agreement with the observed oscillation period of the on-axis accelerating field in Figures \ref{fig:spectra} (a) and (b).

Various strategies can be employed to minimize this effect. First, and most obviously, the mode of the input pulse train should be matched to the lowest-order mode of the channel as closely as possible, which would minimize oscillations associated with $|\delta w|/w_0$. Next the following quantity, which is a measure of the oscillation of the wakefield structure behind the pulse train, should be minimized: $\left|\left(\int_{-\infty}^\infty d\xi(\gamma(\xi)/\omega_w)f^2_\text{acc}(\xi)\right)\big/\left(\int_{-\infty}^\infty d\xi f^2_\text{acc}(\xi)\right)\right|$. For a given wake amplitude $\delta n_a(\xi)/n_{00}$, the two ways to minimize this quantity are: (i) to reduce detuning effects to ensure that the plasma wave zeros are as close to the pulse centroids as possible; and (ii) to increase the modulator parameter $\beta$, which shortens the duration of each pulse in the train, so that they only sample the plasma wave close to its zero-crossings; although we note that this latter strategy also increases SPM, which hastens the onset of depletion effects towards the heads of the pulses. 

\section{Depletion Effects}

We also have to consider that the paraxial description which led to the result in Eq. (\ref{eq:analytic}) breaks down when depletion becomes non-negligible, which becomes important when driving quasi-linear wakes where the dephasing length is a large fraction of the depletion length. 

One caveat to the earlier analysis of the depletion length in Eq. (\ref{eq:L_pd}) is that pulse trains do not deplete uniformly. Instead, trailing pulses, which co-propagate with steeper density gradients, will deplete faster than those near the beginning of the pulse train. We thus expect the pulses towards the rear of the train to have individual depletion lengths up to half as short as in Eq. (\ref{eq:L_pd}), whilst the pulses near the beginning of the train will experience minimal depletion.

The non-uniform depletion effects include group velocity dispersion (GVD) between the pulses, caused by the trailing depleted pulses slowing down more, and detuning spot size oscillations as depleted pulses have shorter spot size oscillation periods due to their shorter Rayleigh lengths. 

\begin{figure}
    \centering
    \subfloat[\label{fig:4a}]{
        \includegraphics[width=1.0\linewidth]{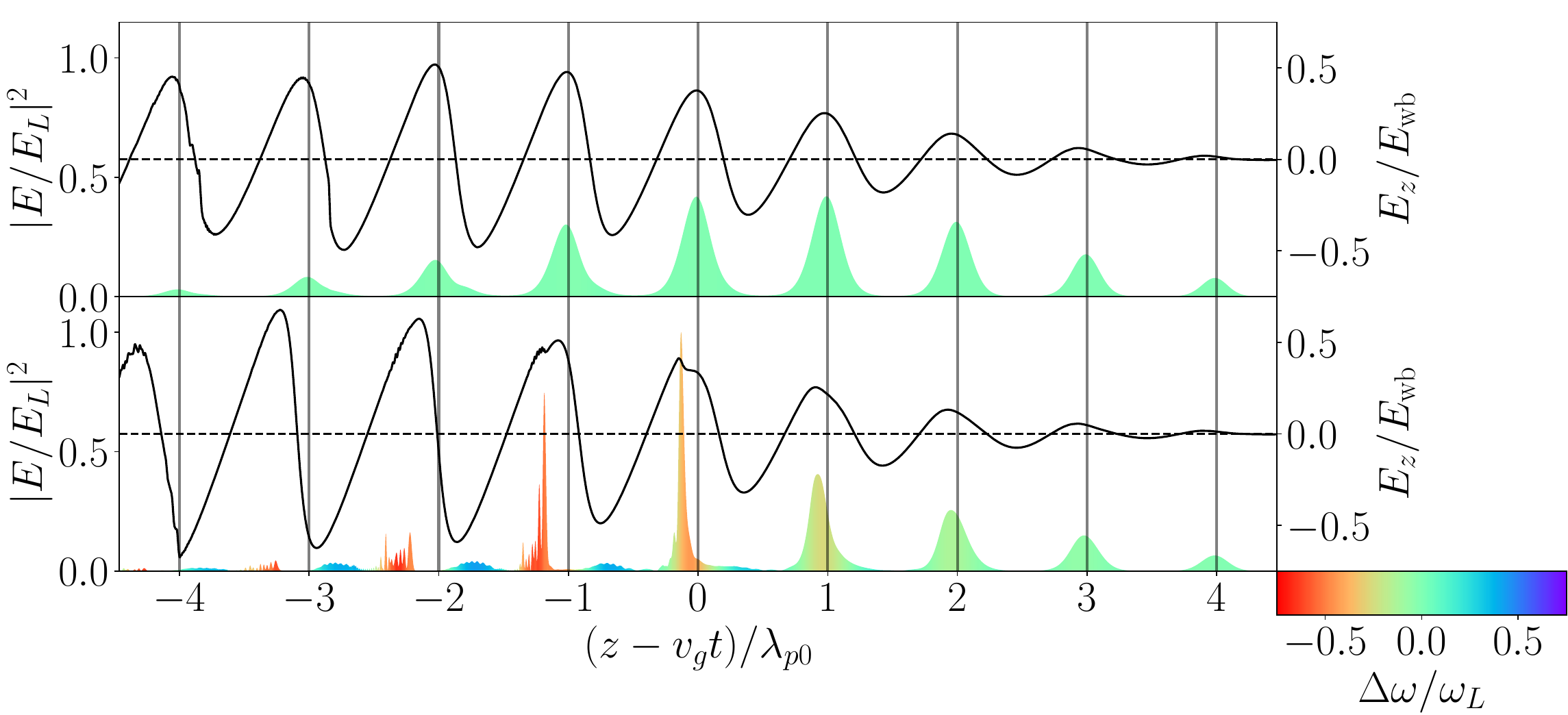}
    }
    \hspace{1cm}
    \subfloat[\label{fig:4b}]{
        \includegraphics[width=1.0\linewidth]{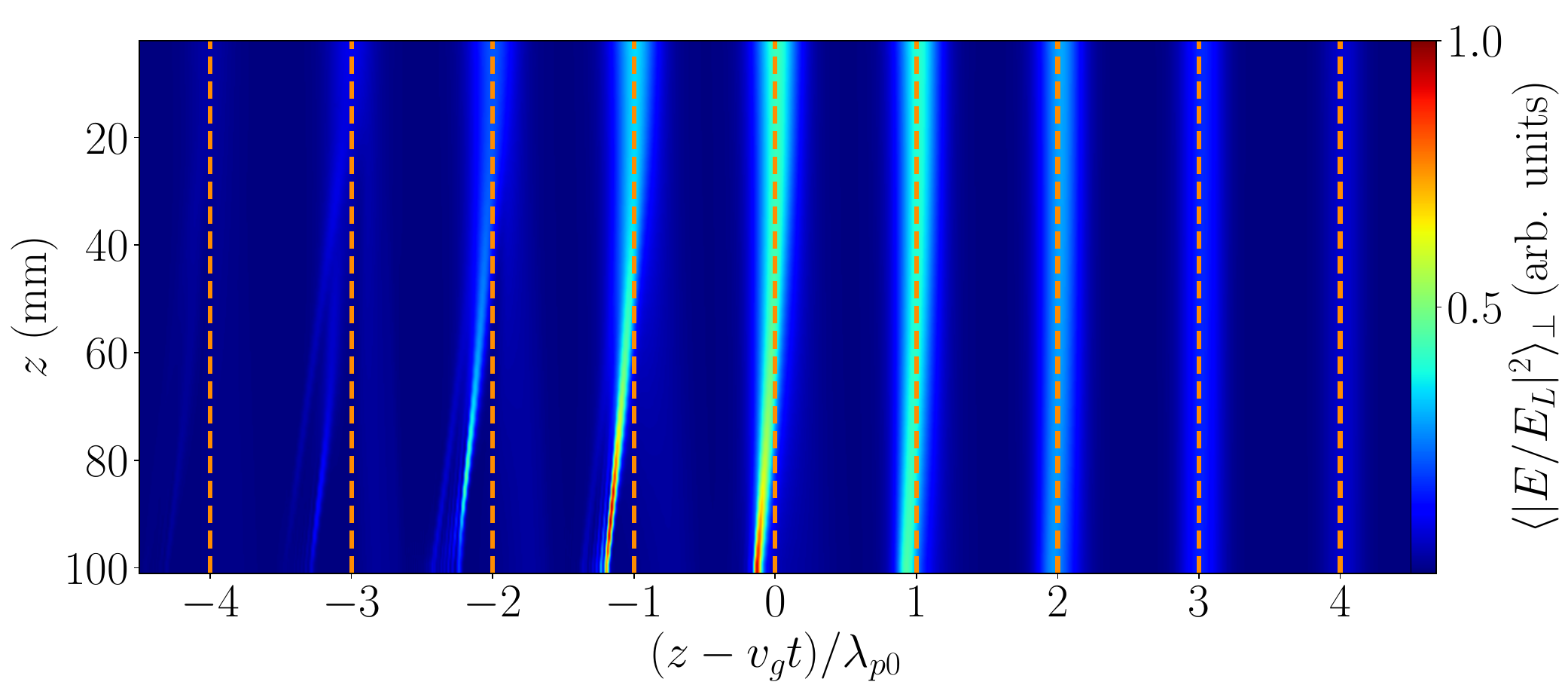}
    }
    \caption{[Color online]. Results from the same PIC simulation presented in Figure \ref{fig:TME}. (a) The on-axis normalized axial electric field $E_z/E_\text{wb}$ (solid, black) and the normalized laser intensity envelope $|E/E_L|^2$ with $E_L=m_e\omega_Lc/e$ and its shift in instantaneous frequency $\Delta\omega/\omega_L$ (color scale) after 2 mm (top) and 100 mm (bottom) of propagation. (b) The evolution of the transverse-averaged intensity envelope $\langle |E/E_L|^2\rangle_\perp$. Note that $|E/E_L|\neq|a|$ at large propagation distances due to the significant red-shifting of the trailing pulses.}
    \label{fig:centroids}
\end{figure}

As demonstrated in Figure \ref{fig:centroids}, the pulse train is initially uniformly spaced and can therefore suffer from Rosenbluth-Liu detuning, though it remains a small effect due to the low number of pulses. However, as the pulse train depletes, the strongly red-shifting trailing pulses fall behind and become resonant with the wake. We also observe a small portion of the heads of each of the trailing pulses strongly blue-shifting and catching up with the train. 

The onset of depletion effects is also expedited by relativistic SPM, as this increases the rate at which the heads of each pulse in the pulse train red-shifts and causes pulse steepening \cite{641309}, which is observed in Figure \ref{fig:centroids}(a). On one hand, depletion causes bunch dephasing to occur sooner than predicted by Eq. (\ref{eq:Ld}) by a reduction in group velocity. On the other hand, the resulting reduction in instantaneous laser frequency increases the ponderomotive force (which scales with $F_p\propto\omega^{-1}$ when holding photon number density constant), causing the wakefield amplitude to gradually increase with propagation distance, similar to the effect observed experimentally by M. J. V. Streeter et al. \cite{PhysRevLett.120.254801}.

\captionsetup[subfloat]{position=top,justification=raggedright,singlelinecheck=false}
\captionsetup[subfloat]{position=bottom,justification=centering,singlelinecheck=true}

\section{Multi-GeV P-MoPA electron accelerator}
\begin{figure*}[tb]
    \centering
        \hspace*{-1.0em}
        \subfloat[\label{fig:5a} $W_\text{drive}=\SI{2.4}{J}$, $R=\SI{30}{\micro m}$, $n_{00}=\SI{2.5e17}{cm^{-3}}$]{
            \includegraphics[width=0.333\linewidth]{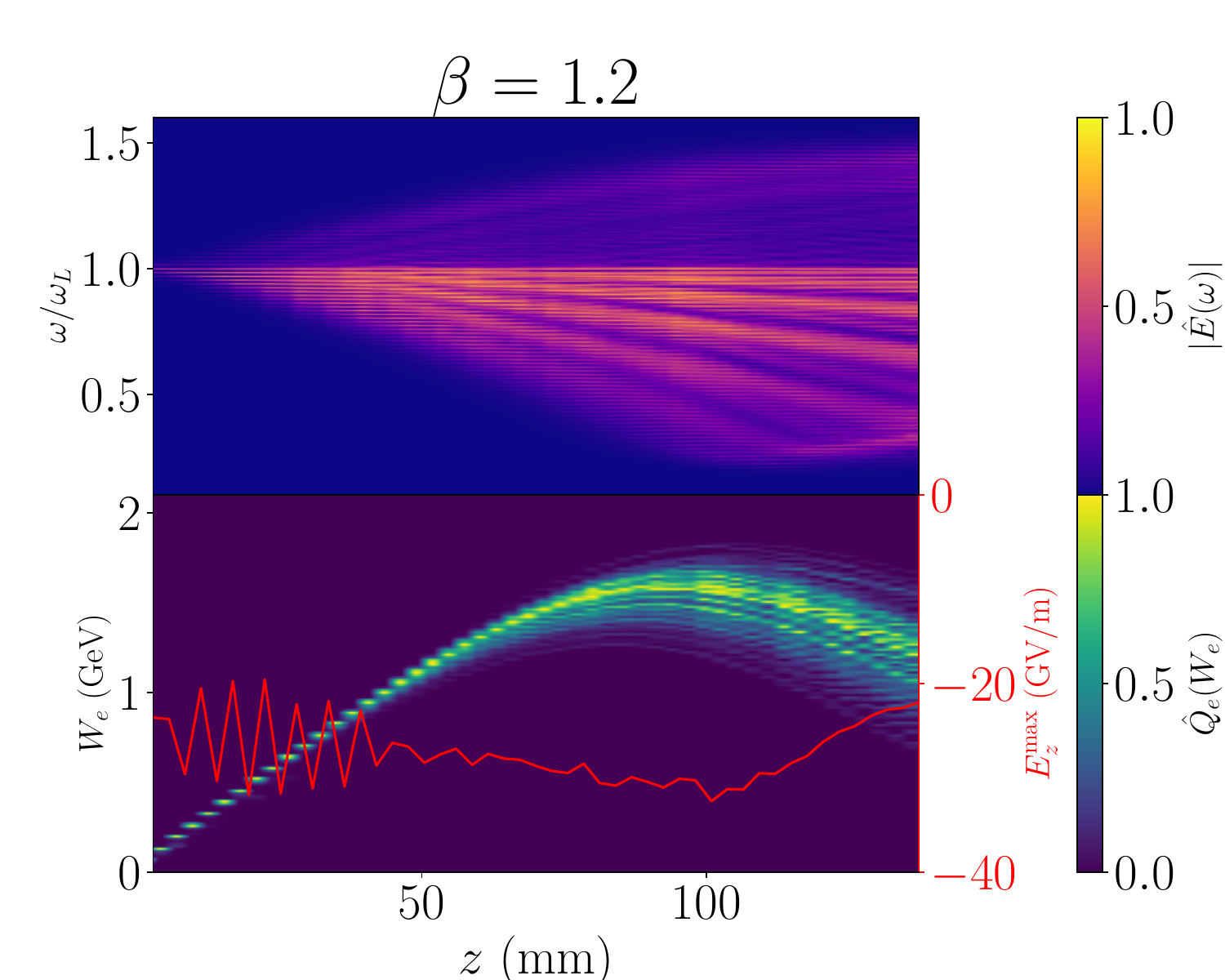}
        }\hspace*{-1.0em}
        \subfloat[\label{fig:5b} $W_\text{drive}=\SI{2.4}{J}$, $R=\SI{30}{\micro m}$, $n_{00}=\SI{2.5e17}{cm^{-3}}$]{
            \includegraphics[width=0.333\linewidth]{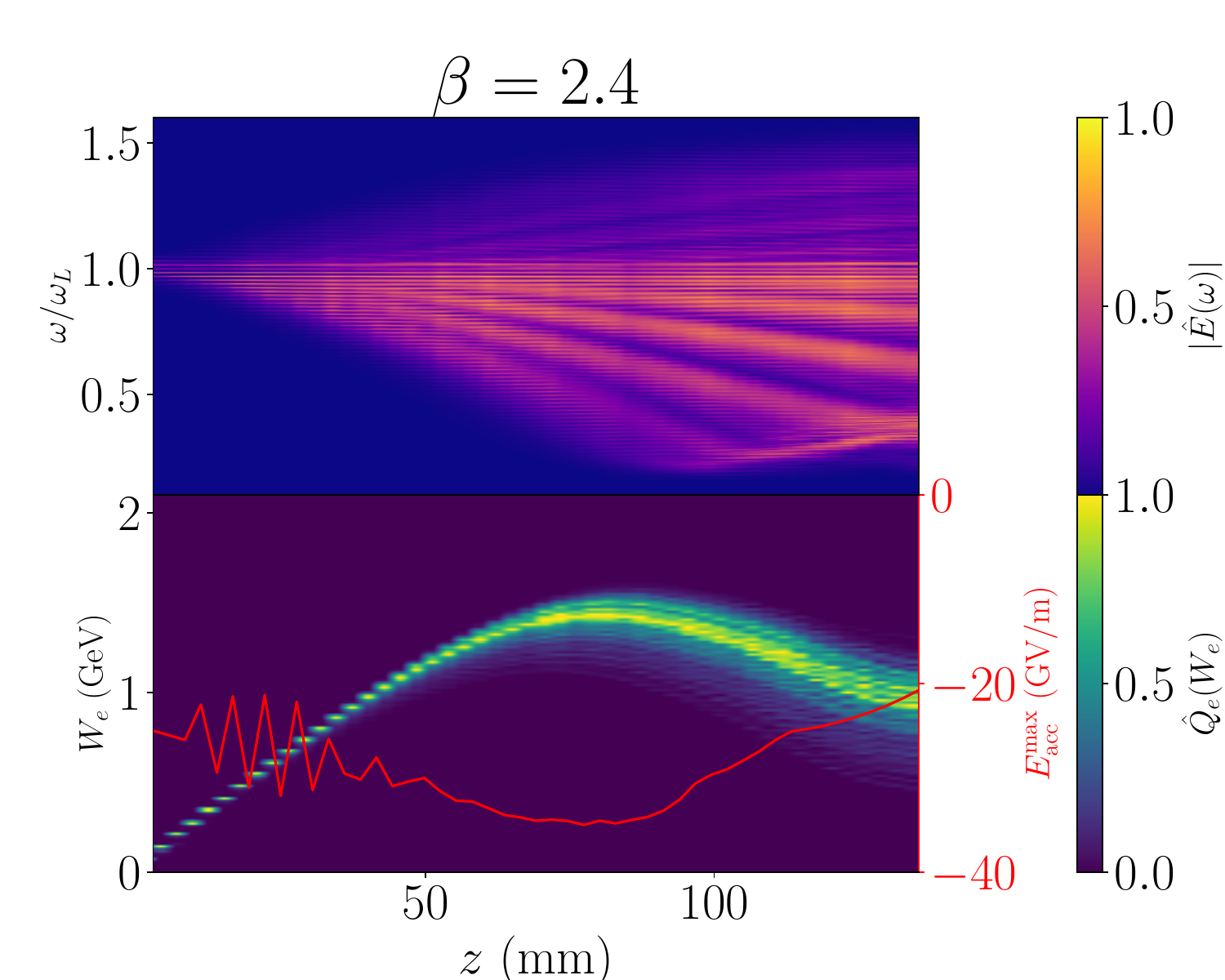}
        }\hspace*{-1.0em}
        \subfloat[\label{fig:5c} $W_\text{drive}=\SI{5.0}{J}$, $R=\SI{36}{\micro m}$, $n_{00}=\SI{1.0e17}{cm^{-3}}$]{
            \includegraphics[width=0.333\linewidth]{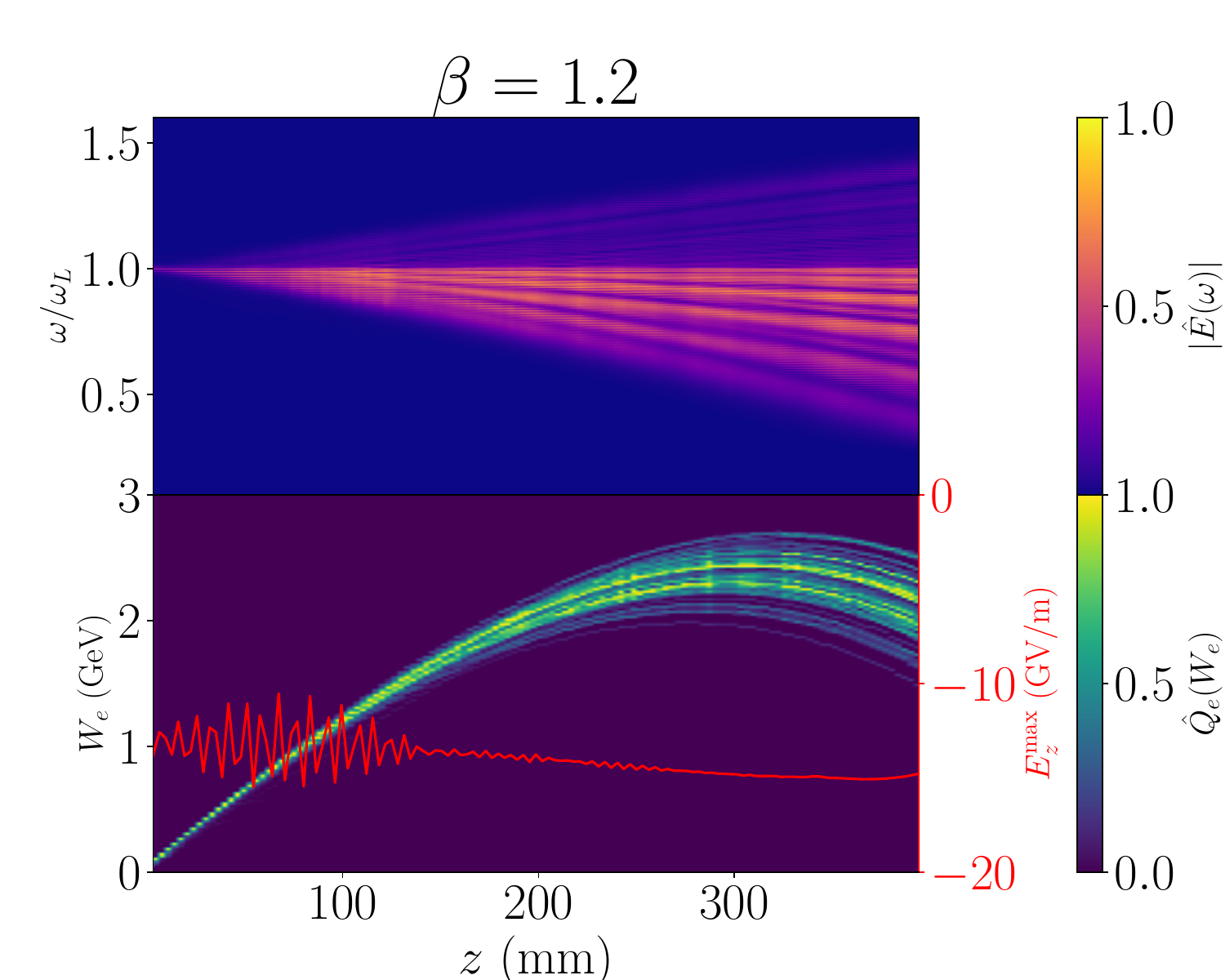}
        }
    \caption{[Color online]. 2D PIC simulation results of accelerator performance at different modulation parameters $\beta$ in a quasi-square channel of the same form used in Fig. \ref{fig:TME}. The top panels display the normalized laser spectrum $|\hat{E}(\omega)|$ evolving with the propagation distance, whilst the bottom panels show the evolution of the normalized injected bunch electron energy spectrum $\hat{Q}_e(W_e)$ and maximum on-axis accelerating field $E_\text{acc}^\text{max}$. Note that the oscillation of the on-axis accelerating field due to spot size oscillations is a continuous sinusoidal motion which disappears beyond $\SI{50}{mm}$ in (a,b) and $\SI{130}{mm}$ in (c), and only appears to be jagged in these plots due to sub-sampling.}
    \label{fig:spectra}
\end{figure*}

Figure \ref{fig:spectra} summarizes the performance of the accelerator stage of a P-MoPA for several different drive laser, and modulator stage parameters. The parameters selected take into account the constraints set by the modulator stage (which sets a limit on $W_\text{drive}$ for a given modulator spot size and plasma density \cite{PhysRevE.108.015204}), relativistic detuning effects, optimizing the wake amplitude (by using a suitable modulator parameter $\beta$) and the evolution with $z$ of the wakefield (caused by SPM and transverse mode excitation). The figures show the evolution of the laser spectra, maximum on-axis accelerating wakefield, and the energy of a test electron bunch. For these 2D PIC simulations, a 1 pC, 35 MeV Gaussian electron bunch was manually placed in phase with the peak accelerating wakefield with dimensions $x_\text{rms} = \SI{4}{\micro m},\,z_\text{rms} = \SI{1.5}{\micro m}$ for $n_{00}=\SI{2.5e17}{cm^{-3}}$. The bunch dimensions were scaled up by a factor of $\sqrt{2.5}$ for the $n_{00}=\SI{1.0e17}{cm^{-3}}$ simulation.

For $n_{00}=\SI{2.5e17}{cm^{-3}}$, the observed propagation distance at which the maximum energy gain is achieved for the $\beta=1.2$ case corresponds well with $L_\text{d}=\SI{93}{mm}$ predicted by Eq. (\ref{eq:Ld}) using $R_\infty^\text{eff}=\SI{46}{\micro m}\Rightarrow w_{0,\text{eff}}=\SI{30}{\micro m}$. The laser-plasma energy transfer is found from the PIC simulations to be $16\%$. According to Eq. (\ref{eq:Ld}), $L_\text{d}$ is independent of $\beta$, but Figure \ref{fig:spectra}(b) shows that increasing $\beta$ to 2.4 causes $L_\text{d}$ to decrease to $\SI{80}{mm}$. The higher value of beta is also found to increase the laser-plasma energy transfer to $19\%$, primarily due to the earlier onset of pump depletion via SPM, as is evident by comparing the laser spectra in Figs. \ref{fig:spectra}(a) and (b). Despite the earlier onset of pump depletion for the higher $\beta$ case, the mean energy gain for both values of $\beta$ is approximately $\sim\SI{1.5}{GeV}$. This value is in good agreement with Eq. (\ref{eq:W_e}), which predicts an energy gain of $\Delta W_e=\SI{1.4}{GeV}$ for a relative wake amplitude of $E_\text{max}/E_\text{wb}=0.5$, consistent with the amplitude observed in PIC simulation shown in Figure \ref{fig:centroids}. 

Figure \ref{fig:spectra}(c) shows the result for a lower density accelerator stage with $n_{00}=\SI{1.0e17}{cm^{-3}}$ and $w_0=\SI{36}{\micro m}$ at $\beta=1.2$. The evolution of the pulse train spectrum, accelerating field, and bunch energy is shown for lengths up to $\SI{400}{mm}$. A bunch of energy around 2.5 GeV is obtained, which is  in good agreement with the value of $\Delta W_e=\SI{2.6}{GeV}$ given by  Eq. (\ref{eq:W_e}) for $E_\text{max}/E_\text{wb}=0.46$ (which is slightly lower than the previous simulations due to the scaled laser energy density $W_\text{drive}/w_0^2\lambda_{p0}$ being $8.5\%$ smaller). Figure \ref{fig:spectra}(c) shows that dephasing occurs at $z = L_d=\SI{300}{mm}$, which is close to the value of $\SI{296}{mm}$ predicted from Eq. (\ref{eq:Ld}). It is clear from these simulations that P-MoPAs operating at multi-GeV energy gains should be possible.

All three simulations shown in Figure \ref{fig:spectra} exhibit the expected oscillation in peak on-axis accelerating field due to transverse oscillations of the pulses in the train, as discussed in the previous section and observed in Figure. \ref{fig:TME}. In addition, Figure \ref{fig:spectra} shows that the accelerating field initially grows in magnitude with propagation. This increase arises from: (i) SPM-induced pulse steepening; (ii) the effect shown in Figure \ref{fig:centroids}, whereby greater red-shifting of the trailing pulses in the train causes the pulse separation to increase towards the back of the train, partially circumventing Rosenbluth-Liu detuning; and (iii) an increase in the ponderomotive force exerted by the trailing pulses, which arises from their greater red-shift.

\section{Conclusion}

We have studied the operation of the accelerator stage in a P-MoPA using both the paraxial wave equation in the weakly relativistic limit and 2D PIC simulations.

The ability to adjust the modulator parameter $\beta$ was found to bring several advantages. First, unlike in the plasma beat-wave accelerator, it allows the temporal envelope of pulses within the pulse train to be controlled. For the optimal value of  $\beta=1.43$, the wake amplitude was found to be $72\%$ larger than that generated in a PBWA with the same total drive energy. Second, adjusting $\beta$ can minimize transverse mode excitation, and hence oscillation of the wakefield structure, which is likely in turn to minimize increases in the energy spread and emittance of an injected electron bunch. 

Rosenbluth-Liu detuning was found to be much less important in a P-MoPA than it is in a PBWA, for two reasons. First, our analysis shows that in the linear and quasi-linear regimes Rosenbluth-Liu detuning would not occur appreciably if the number of pulses in the P-MoPA train was less than $\sim 30$. Second, in a P-MoPA operating in the quasi-linear regime, this detuning is at least partially counteracted by the increase in pulse spacing towards the back of the train caused by increased red-shifting.

We presented an analysis of transverse mode oscillations within the paraxial approximation, which was found to be in good agreement with 2D PIC simulations. PIC simulations for several combinations of laser and plasma parameters showed good agreement with expressions derived for the spot size oscillation frequency, pump depletion, dephasing length and energy gain. These simulations demonstrate that particle energy gains of several GeV should be possible in a P-MoPA.

We note that in this study we ignored the effects of the accelerated bunch on the wakefield. In addition to decreasing the acceleration gradient --- which has the compensating advantage of providing a means to reduce the bunch energy spread --- beamloading effects could introduce a coupling between the pulse train and the injected bunch dynamics. This will be a subject for future research.

The work presented in this paper confirms that P-MoPAs driven by few-joule, picosecond pulses, such as those provided by high-repetition-rate thin-disk lasers, have the potential to accelerate electron bunches to multi-GeV energies at pulse repetition rates in the kilohertz range.

\acknowledgements

This work was supported by the UK Engineering and Physical Sciences Research Council (EPSRC) (Grant No.\ EP/V006797/1), the UK Science and Technologies Facilities Council (Grant No.\ ST/V001655/1]), InnovateUK (Grant No.\ 10059294), United Kingdom Research and Innovation (UKRI) ARCHER2 Pioneer Projects (ARCHER2 PR17125) \cite{archer2} and the Ken and Veronica Tregidgo Scholarship in Atomic and Laser Physics. This publication arises from research funded by the John Fell Oxford University Press Research Fund. This research used the open-source particle-in-cell code WarpX \cite{WarpX} \url{https://github.com/ECP-WarpX/WarpX}, primarily funded by the US DOE Exascale Computing Project. Primary WarpX contributors are with LBNL, LLNL, CEA-LIDYL, SLAC, DESY, CERN, and Modern Electron. We acknowledge all WarpX contributors.

This research was funded in whole, or in part, by EPSRC and STFC, which are Plan S funders. For the purpose of Open Access, the author has applied a CC BY public copyright licence to any Author Accepted Manuscript version arising from this submission.

The input decks used for the PIC simulations presented in this paper are available at \url{https://doi.org/10.5281/zenodo.10061009}

\bibliography{references}

\end{document}